\RequirePackage{fix-cm}
\documentclass[smallextended]{svjour3}       
\smartqed  
\usepackage{graphicx}
\usepackage{amsmath}
\begin{document}

\title{The Pointing System of the \textit{Herschel}\footnote{Herschel is an ESA space observatory with science instruments provided by European-led Principal Investigator consortia and with important participation from NASA.} Space Observatory
}
\subtitle{Description, Calibration, Performance and Improvements}

\titlerunning{Herschel's Pointing: Calibration, Performance and Improvements}        

\author{Miguel S\'anchez-Portal \and
Anthony Marston \and
Bruno Altieri \and
Herv\'e Aussel \and
Helmut Feuchtgruber \and
Ulrich Klaas \and
Hendrik Linz \and
Dieter Lutz \and
Bruno Mer\'{\i}n \and
Thomas M\"uller \and
Markus Nielbock \and
Marc Oort \and
G\"oran Pilbratt \and
Micha Schmidt \and
Craig Stephenson \and
Mark Tuttlebee \and
The Herschel Pointing Working Group
}


\institute{Miguel S\'anchez-Portal \at
              ISDEFE for ESA, Herschel Science Centre, European Space Astronomy Centre (ESAC), Villanueva de la Ca\~nada, E-28691 Madrid, Spain \\
              Tel.: +34-91-8131349 \email{miguel.sanchez@sciops.esa.int}           
\and 
Anthony Marston \and Bruno Altieri \and Bruno Mer\'{\i}n \at
              Herschel Science Centre, ESAC/ESA, Villanueva de la Ca\~nada, E-28691 Madrid, Spain
\and 
Herv\'e Aussel \at
CNRS/Service d'Astrophysique,  Bat. 709, 91191 Gif-sur-Yvette, France
\and
Helmut Feuchtgruber \and Dieter Lutz \and Thomas M\"uller \at
Max-Planck-Institut f\"ur extraterrestrische Physik, Postfach 1312, 85741 Garching, Germany
\and
Ulrich Klaas \and Hendrik Linz \and Markus Nielbock \at
Max-Planck-Institut f\"ur Astronomie, K\"onigstuhl 17, D-69117 Heidelberg, Germany
\and
Marc Oort \at
Dutch Space B.V., Leiden, The Netherlands.
\and
G\"oran Pilbratt \at
ESA Astrophysics Missions Div./Research and Scientific Support Dept., ESTEC/SRE-SA,  Keplerlaan 1, NL-2201 AZ Noordwijk, The Netherlands
\and
Micha Schmidt \at
European Space Operations Centre (ESOC)/ESA, Robert-Bosch Strasse 5, D-64293 Darmstadt, Germany
\and
Craig Stephenson \at
              Telespazio Vega for ESA, Herschel Science Centre, ESAC, Villanueva de la Ca\~nada, E-28691 Madrid, Spain \\
\and 
Mark Tuttlebee \at
SCISYS for ESA, ESOC, Robert-Bosch Strasse 5, D-64293 Darmstadt, Germany
}
\date{Received: date / Accepted: date}

\maketitle

\begin{abstract}
We present the activities carried out to calibrate and characterise the performance of the elements of attitude control and measurement on board the Herschel spacecraft. The main calibration parameters and the evolution of the indicators of the pointing performance are described, from the initial values derived from the observations carried out in the performance verification phase to those attained in the last year and half of mission, an absolute pointing error around or even below 1\,arcsec, a spatial relative pointing error of some 1\,arcsec and a pointing stability below 0.2\,arsec. The actions carried out at the ground segment to improve the spacecraft pointing measurements are outlined. On-going and future developments towards a final refinement of the Herschel astrometry are also summarised. A brief description of the different components of the attitude control and measurement system (both in the space and in the ground segments) is also given for reference.  We stress the importance of the cooperation between the different actors (scientists, flight dynamics and systems engineers, attitude control and measurement hardware designers, star-tracker manufacturers, etc.) to attain the final level of performance.

\end{abstract}

\section{Introduction}
\label{intro}
\begin{figure}
\centering
  \includegraphics[scale=0.6]{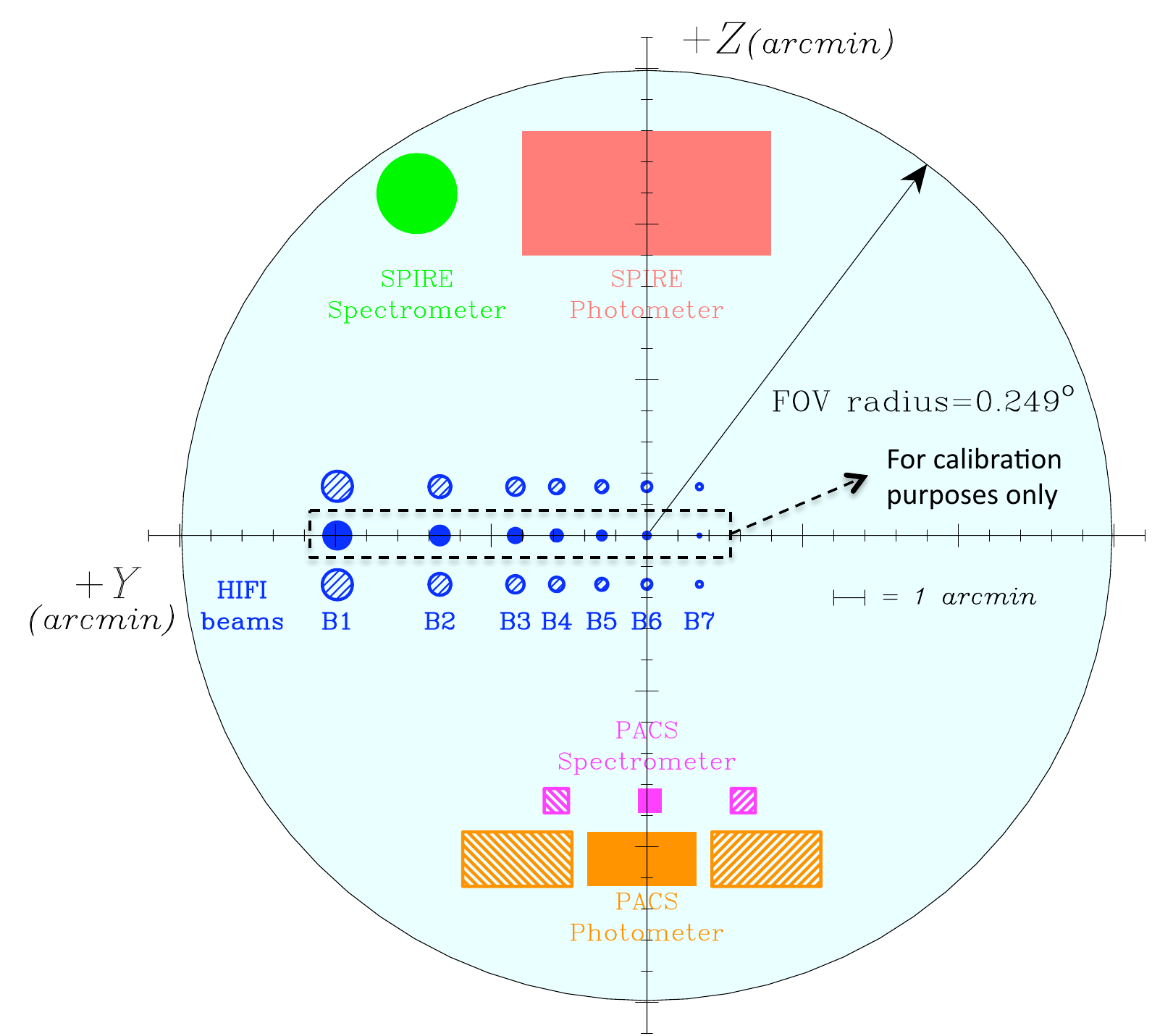}
\caption{Location of the different instruments' apertures within the Herschel FoV}
\label{herschel_fov}       
\end{figure}

The Herschel Space Observatory \cite{herschel}  is the fourth ``cornerstone'' mission in the ESA Horizon 2000 science programme. It has been the first large aperture space far--infrared (FIR)/sub--mm observatory, performing  photometry and spectroscopy in
the 55--672\,$\mu$m wavelength range. The observatory consisted  of a telescope with a large (3.5\,m) monolithic low emissivity passively cooled mirror
and three focal plane science instruments enclosed in a large HeII cryostat: The PACS instrument \cite{pacs}, consisting of \textit{(a)} a dual-band photometer, comprising a  ``blue photometer'' with two filters mounted in a wheel, allowing to select a range centred at either 70\,$\mu$m  or  100\,$\mu$m and a ``red photometer''  covering a spectral range centred at 160\,$\mu$m; and  \textit{(b)} an integral field spectrometer in the range \mbox{57-210 $\mu$m}. The photometric bands covered the same field-of-view (FoV), while the FoV of the spectrometer was offset from the photometer. The SPIRE instrument \cite{spire} consisted of a three-band imaging photometer and an imaging Fourier Transform Spectrometer. As for the PACS instrument, the photometer and spectrometer did not operate simultaneously.  The three photometer arrays  provided broad-band photometry  
in bands centred at 250, 350 and 500~$\mu$m with a common FoV, and the spectrometer was also offset from the photometer. Finally, the HIFI instrument  \cite{hifi} was a high-resolution heterodyne spectrometer with a continuous coverage 
from 480 to 1250~GHz in five bands, plus two bands providing a coverage of the 1410--1910~GHz range.  Both polarisations of the astronomical signal were detected for maximum sensitivity. To cover a wide frequency range with high sensitivity, HIFI was designed to have 7 mixer
bands and 14 Local Oscillator 
sub-bands (two frequency sections per band). Figure \ref{herschel_fov} shows the position of the instruments' apertures within the Herschel FoV.

The Herschel spacecraft (S/C) had a modular design, comprising the Extended Payload Module (EPLM) and the Service Module (SVM). The EPLM consisted of the payload module described above, the sunshade and solar array and payload associated equipment. The SVM housed ``warm'' payload electronics and provided services such as power, attitude and orbit control, the on-board data handling and command execution, communications, and safety. 

The Herschel S/C was stabilised in three-axes. The attitude was commanded by means of reaction wheels and controlled using star trackers and gyroscopes, as described in Section \ref{sec:acms}. The S/C pointing modes were based either on stare pointings (fine pointing mode) or moving pointings at constant rate (line scan mode). Raster maps are ``grids'' of stare pointings at regular spacings; scan maps are sequences of line scans at regular spacing. Allowed scan angular speed ranged from 0.1\,arcsec/s to 1\,arcmin/s. In addition, the Herschel spacecraft could track moving Solar System targets at rates up to 10\,arcsec/min. The S/C  slewed between different pointing positions at rates up to 7\,arcmin/s.

The pointing calibration tasks were aimed, on the one hand, to provide an accurate alignment of the different fields of view of each instrument and instrument mode described above with respect to the telescope and the S/C attitude reference frame, and on the other, to characterise the performance of the various pointing modes. This paper describes these activities and is structured as follows: in Section \ref{sec:acms}, we discuss the basic elements of the Attitude Control and Measurement System (ACMS) 
and the route of telemetry (TM) data from the spacecraft to the Herschel Science Centre (HSC) at the European Space Astronomy Centre (ESAC). Section \ref{sec:products} is devoted to the pointing-related products generated at the HSC: the pointing product, the ACMS TM product and the Spacecraft/Instrument Alignment Matrices (SIAM) product. Section \ref{sec:calib_params} deals with the main calibration parameters of the Herschel spacecraft pointing system and historical description of the evolution of the indicators of its performance, that in general met or surpassed the requirements set.  However, given a clear scientific motivation for further improvement (see Section \ref{sec:calib_params}), an outstanding effort has been done and is still being invested: in Section \ref{sec:pointing_improvement}, the work carried out within the ground segment centres to improve the S/C pointing measurements is described. On-going and future developments towards a final refinement of the Herschel astrometry are also described.

\section{Basic elements and operation of the ACMS}
\label{sec:acms}
\begin{figure}
  \includegraphics[scale=0.8]{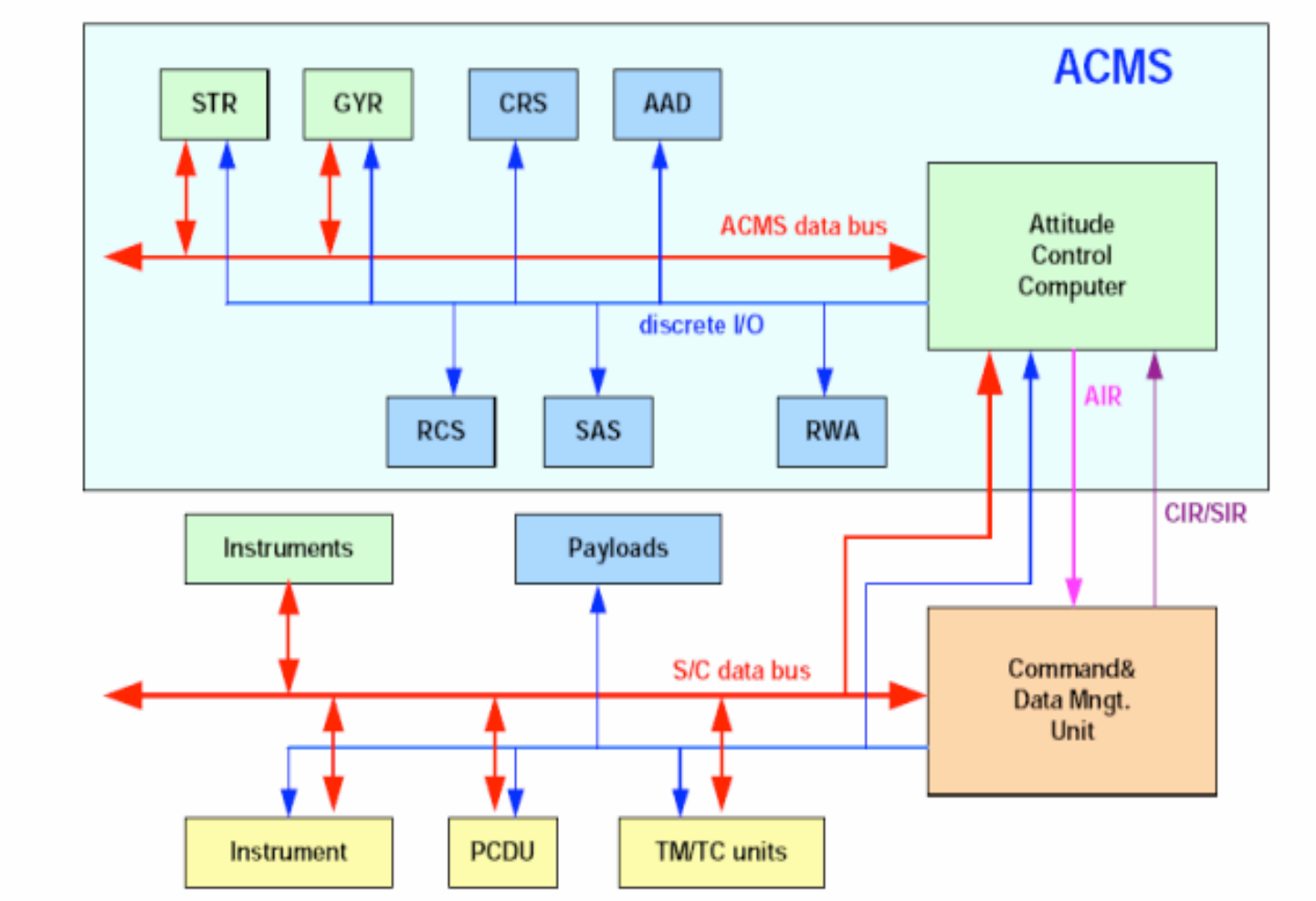}
\caption{ACMS block diagram (reproduced from \cite{pointing_calibration_plan}). The individual elements depicted and
   their acronyms are described in the text.}
\label{acms_blocks}       
\end{figure}

The control of the positioning of the Herschel spacecraft (S/C) was carried out by its Attitude Control and Measurement System (ACMS). The main components were \cite{acms}: {\em (i)} the attitude control computer (ACC); {\em (ii)} gyroscopes (GYR; four units); {\em (iii)} star trackers (STR): two units, with boresights approximately opposite to that of the telescope; {\em (iv)} the assembly of four reaction wheels (RWA) to perform pointing manoeuvres; {\em (v)} the Reaction control system (RCS), i.e. the set of thrusters used for orbit manoeuvres and reaction wheel biasing; {\em (vi)} the sun acquisition sensors (SAS); {\em (vii)} the coarse rate sensors (CRS); and {\em (viii)} the attitude anomaly detectors (AAD). The different elements of the ACMS are depicted in Fig. \ref{acms_blocks} and described in \cite{acms}.

%


The star trackers (STR) were small telescopes with CCD detectors whose main function was the determination of the S/C attitude. The Herschel S/C had two (cold) redundant autonomous units built by Selex Galileo S.p.A. \cite{str_um}. The STR telescope had a focal length of 30\,mm and an f/number of 1.2, yielding a large FoV of 16.4\,$\times$\,16.4\,deg$^2$. The CCD detector used was the Atmel TH78900, with 512\,$\times$\,512\,pixels of 17\,$\mu$m. The plate scale was therefore 116.9\,arcsec/pixel (the CCD area is slightly larger than the FoV). The Herschel STRs had the ability to determine the inertial position from a ``lost in space'' situation, or tracking from a known attitude. To this end, an on-board star catalogue, based on Hipparcos, was used. The criteria applied to build this catalogue included a specific range of brightness, a \mbox{B--V} colour threshold, a low variability tolerance, reduced proper motion and  absence of close companions or neighbours \cite{str_catalogue}. The on-board catalogue
contained 3599 stars, out of which originally 3047 (85\%, but cf. Section \ref{subsec:od1011_to_eol}) were enabled for tracking purposes. A minimum of 3 stars should be acquired within the FoV. Due to hardware limitations, a maximum of  9 stars could be simultaneously used for attitude determination. Nevertheless, there was an enhanced performance mode called ``interlacing function'', only applicable if $\geq$\,15 trackable stars were located within the FoV. In this mode, the STR determined the attitude using two different sets of stars from two consecutive frames, (i.e. each set of stars were sampled at half the nominal 4\,Hz sampling frequency). In this way, up to 18 stars could be used for attitude determination. Due to the large weight of the STR measures in the attitude estimation, the highly undersampled STR FoV was the largest contributor to the absolute pointing error (APE). On top of that, the position-dependent STR bias was caused by flat-fielding effects and optical distortions. On average, the contribution to the APE was  0.8\,$\times$\,$\sqrt{2}$\,arcsec.

\begin{figure}
\hspace{-1cm}
  \includegraphics[scale=1.0]{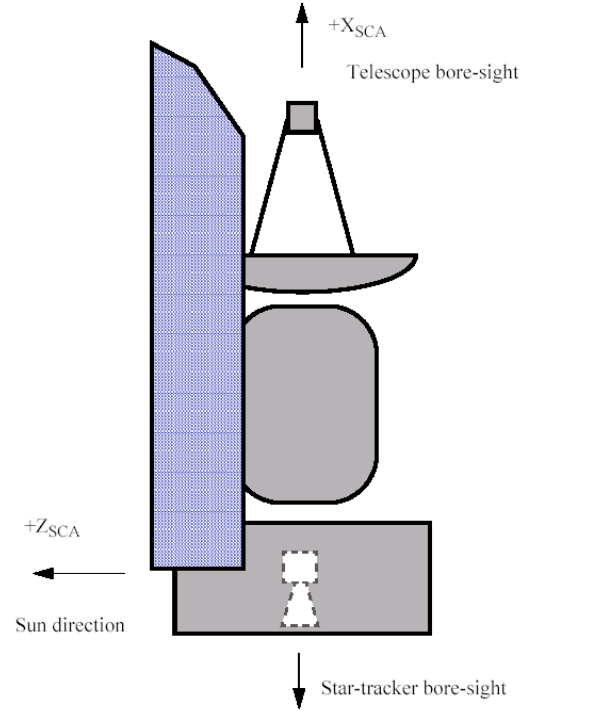}
  \hspace{-0.7cm}
  \includegraphics[scale=0.5]{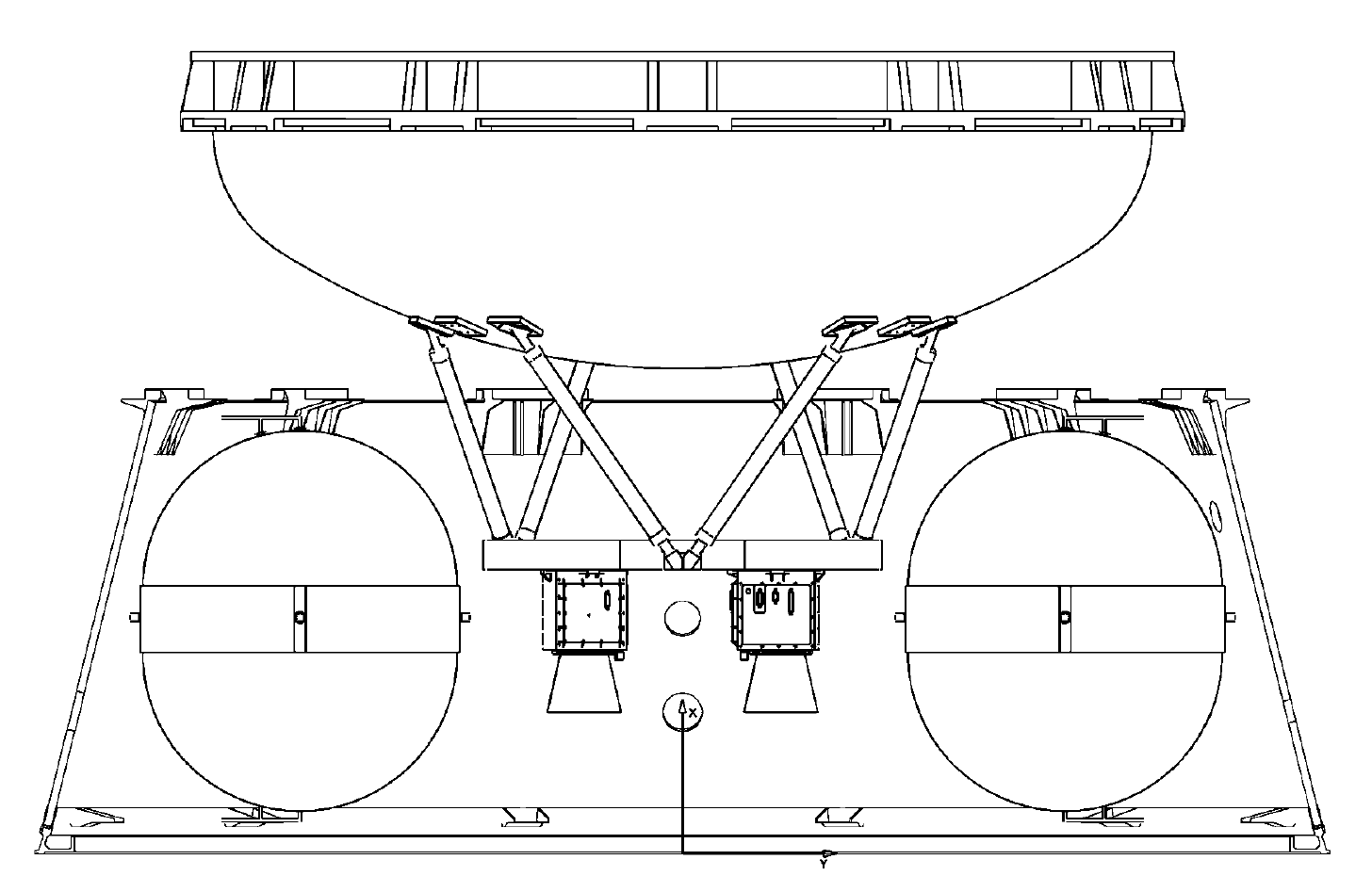}
\caption{Left: Herschel spacecraft axes, telescope and boresight of the STRs (reproduced from \cite{pointing_calibration_plan}). Right: diagram of the two redundant STR mountings below the CVV. The STRs are shown as the two identical, small square boxes with conical baffles (reproduced from \cite{elfving_rasmussen})}
\label{str_sketch}       
\end{figure}

The mounting of the STRs with respect to the S/C and telescope viewing direction is shown in Fig. \ref{str_sketch} (left). Note that the STRs pointed approximately at 180$^{\circ}$ w.r.t. the telescope boresight ($+X_{SCA}$; see definition in Section \ref{sec:products} below). Both STRs were mounted on a carbon fibre platform suspended by struts from the lower dome of the cryostat vacuum vessel (CVV) as shown in Fig. \ref{str_sketch} (right). The struts were partly made of glass fibre, to minimise the conducted heat input to the CVV, and partly of carbon fibre, to minimise the thermoelastic distortions \cite{elfving_rasmussen}. 

Only one STR could be operative at a given time, the other unit kept as backup. The possibility of using both STRs simultaneously to ease the calibration of their mutual misalignment was ruled out due to the alteration of the pointing due thermoelastic distortion caused by  the excessive heat dissipated. 

The Herschel S/C used hemispherical resonating gyroscopes. These consist of a thin-walled quartz shell that is energized by an electrical field to produce a vibration pattern within itself. The gyroscopes were used to sense and respond to changes in the inertial orientation of its spin axis. 
Rate/rate-integrating gyroscopes provided high-precision measures of the S/C angular speed. The Herschel's ACMS was provided with four gyroscopes mounted in a tetrahedral configuration. The four gyroscopes were hot-redundant. The fourth gyroscope was not used for control, but served to detect an inconsistency in the output of the other three.

Estimates of the satellite attitude quaternion\footnote{
Quaternions are a single example of a more  general class of hypercomplex numbers. By analogy with the complex numbers,
a quaternion can also be written as a linear combination of a real and three imaginary components. It is easy to represent the concatenation of rotations in quaternion format, constituting a very convenient and compact means of representing S/C attitudes. It is outside the scope of this paper to give even an introduction to quaternions. A good introduction to this topic is given in \cite{Wertz}.} and inertial rate vector were obtained combining data from GYR and STR. The estimator was based on a ``model replacement approach'' in which a dynamical model for the evolution of the quaternion states was replaced by integration of the kinematics’ equations of motion, driven by angular rate measurements from the gyro. In the nominal estimator the state (attitude quaternion, rate estimate and GYR bias) was propagated from the previous estimate using the gyro data. This was then combined with the latest STR measurement to obtain an updated state estimate which was subsequently used for control. The main reason for the inclusion of GYR data in the model is that the STR measurements were delayed by 1.5 cycles due to integration and internal processing. Finally, the actuation was performed by means of the RWA. 

The attitude information derived by the ACMS was down-linked at a rate of 4\,Hz\footnote{The actual packet rate was 1\,Hz but many ACMS parameters were super-commutated (four samples per packet).} when the S/C is in normal science mode (SCM) operation. The telemetry (TM) was processed at the Mission Operations Centre (MOC). The Flight Dynamics Team produced the so-called ``attitude  history file'' (AHF) per operations day (OD) basis. The AHF packed the ACMS SCM TM generated at 4\,Hz for slew, point and line scan operations.  It was an ASCII-formatted file that included the 
on-board time, commanded attitude, ACMS-processed attitude (the so-called filtered attitude), three-axes X, Y and Z calibrated angular rates, S/C solar aspect angle (SAA), measurement quality flags etc. The AHF was transferred automatically to the HSC within 8 hours of the end of the OD. It triggered the start of the processing of auxiliary products (e.g. housekeeping telemetry, radiation monitory data, orbit data, etc.) at the HSC.

\section{Pointing products created at the HSC}
\label{sec:products}

The information produced by the ACMS was processed (and is still re-processed) by the HSC and stored in ``products''. These data structures can be accessed from the observation context (a sort of container of the scientific and housekeeping data belonging to a given observation).  The pointing-related products belong to the ``auxiliary products'' category. Three classes have been defined: pointing product, ACMS TM product and spacecraft/instrument alignment matrices' (SIAM) product.  

The pointing product is the most important  of these data structures. It packs the information contained in the AHF, combined with up-link information from the planned observation sequence (POS) file. 
The pointing product consists of sequences of table data sets (analogous to FITS binary tables) including: on-board time, commanded attitude quaternion, different estimates of the attained attitude quaternion, attitude error estimates, number of tracked stars, etc. It is produced per OD, but sliced per observation. The detailed contents of the product can be found in Table \ref{pointingProd}.

The ACMS TM product contains all the relevant information required to re-estimate the S/C attitude from scratch. Sliced per observation, uses essential TM, SCM TM and diagnostics TM. It includes many ACMS database parameters as metadata entries (analogous to the FITS header's keywords). Product's data records include estimated attitude quaternion, STR quaternion, calibrated gyro rates, IDs, positions, magnitudes and colours of tracked stars, GYR calibration parameters etc.  

The spacecraft/instrument alignment matrices’ (SIAM) product contains 3\,$\times$\,3 proper orthogonal rotation matrices DCM$_{INS-ACA}$ converting the coordinates of a vector in the Herschel spacecraft reference frame (spacecraft control axes, SCA; also known as attitude control axes, ACA) into a given instrument reference frame (INS), i.e.:

\begin{equation}
\begin{gathered}
\mathbf{u}_{INS} = DCM_{INS-SCA} \cdot  \mathbf{u}_{SCA} \\
\mathbf{u}_{SCA} = DCM_{INS-SCA}^T \cdot  \mathbf{u}_{INS}
\end{gathered}
\end{equation}

The product contains one rotation matrix per instrument (see a brief description of the instruments in Section in \ref{intro}), i.e. 2 matrices for PACS (one for the photometer aperture and one for the spectrometer one), 56 matrices for HIFI (the apertures are defined for the 7 bands, 2 sub-bands per band, 2 polarisations and 2 chopper positions) and 75 matrices for SPIRE (the SPIRE apertures are defined at pixel level, 17 for the photometer and 57 for the spectrometer, plus an additional aperture for the so-called SPIRE/PACS parallel mode; this aperture is placed in the midpoint between the PACS and SPIRE photometers, approximately in the centre of the telescope FoV. In normal scientific operation, only the central photometer and spectrometer apertures, plus the parallel mode one, are used). The alignment matrices were determined by means of dedicated calibration observations, mainly during performance verification phase (see Section \ref{sec:calib_params}).
For astrometrical purposes, there are two valid SIAM files across the whole mission: 0122\_0001 (OD\,$<$\,320) and 0341\_0001(OD\,$\geq$\,320). This is due to the change in the STR CCD reference temperature that took place in OD320 and will be thoroughly explained in the next section.

\section{Main pointing calibration parameters and evolution of the pointing performance}
\label{sec:calib_params}

The main pointing calibration activities were carried out during commissioning (CoP; some 60 ODs) and performance verification (PVP; approximately 90 ODs) phases using the scientific instruments. These calibrations were aimed, on the one hand, to accurately determine the location of the different instrument apertures (i.e. computation of the SIAMs), and on the other hand, to derive the main figures of merit characterising the pointing performance. These are according to \cite{esa_pointing_hb} the Absolute Pointing Error (APE), defined as the angular separation between the desired direction and the instantaneous actual direction; the Pointing Drift Error (PDE), defined as the angular separation between the average pointing direction over some interval (nominally 55\,min) and a similar average at a later time (nominally after 24\,hours), is a measure of the long-term stability of the S/C pointing, while the Relative Pointing Error (RPE, also referred to as pointing stability) is the angular separation between the instantaneous pointing direction and the short-time average pointing direction at a given time period (fixed at 60\,s); finally, the Spatial Relative Pointing Error (SRPE) is defined as the angular separation between the average orientation of the satellite fixed axis and a pointing reference axis which is defined relative to an initial reference direction: this measurement is important to characterise  the accuracy of the relative astrometry in a map comprising several pointings (e.g. a raster or scan map).  

The pointing error specifications are expressed as half-cone angles of the line-of-sight (LoS) and half-sector angles around the LoS. They are specified at a temporal probability level of 68\%, which implies that the error will be less than the requirement for 68\% of the time for each pointing direction. For Herschel, it was required that the APE in stare pointings  was 3.7\,arcsec (LoS) and 3.0\,arcmin (around LoS).  The PDE requirement (measured within a 24\,h period) is 1.20\,arcsec (LoS) and 3.0\,arcmin (around LoS). The RPE requirement (for a period of 60\,s)  is 0.30\,arcsec (LoS) and 1.5\,arcmin (around LoS).  Finally, regarding the spatial relative pointing accuracy, it was required that, in consecutive pointings within a 4$^{\circ}$\,$\times$\,4$^{\circ}$ spherical area, the SRPE of all pointings following the initial pointing, as referred to the average pointing direction of the first pointing shall be less than 1\,arcsec (LoS).  Table \ref{req_pred} summarises the initial requirements for the main pointing figures of merit as given in the early mission definition phase and their predicted performance from pre-launch testing campaigns. It was anticipated a non-compliance of the SRPE (2.24\,arcsec predicted vs. 1.0\,arcsec required) but the rest of the requirements were clearly met or exceeded. As will be shown below, the in-orbit performance of the Herschel S/C confirmed these predictions. However, beyond the early defined requirements shown in Table \ref{req_pred} (issued when the final characteristics of the instruments were only barely known), and given Herschel instruments' beams as small as 5\,arcsec FWHM at the shortest wavelengths, there was a clear scientific motivation for further improvement, for better centring of spectroscopic observations on source, as well as for better a posteriori reconstruction of the actual pointing in imaging and spectroscopy.

\begin{table}[htdp]
\caption{Initial requirements and pre-flight predicted performance of key pointing accuracy indicators. Goal conditions assume 18 stars available for guidance within the STR (i.e. the maximum possible number).}
\label{req_pred}
\begin{tabular}{lcccc}
\hline\noalign{\smallskip}
& \multicolumn{2}{c}{Baseline (arcsec)} & \multicolumn{2}{c}{Goals (arcsec)}\\
Parameter & Requirement & Prediction & Requirement & Prediction \\
APE point & 3.7 & 2.24 & 1.5 & 1.34\\
APE scan & 3.7 & 2.36 & 1.5 & 1.42\\
SRPE & 1.0 & 2.21 & 1.0 & 1.52\\
RPE & 0.3 & 0.24 & -- & --\\  
PDE & 1.2 & 088 & -- & --\\
\noalign{\smallskip}\hline
\end{tabular}
\end{table}

The SIAM matrices were initially defined by means of accurate measurements performed on-ground during the assembly, integration and verification phase (AIV). In flight, new SIAMs were derived by planning and executing dedicated calibration observations with the PACS, SPIRE and HIFI instruments, as described in \cite{pointing_calibration_plan} and \cite{pointing_calibration_report}. The first calibration (serving as reference for the initial bulk shift of all instruments' apertures with respect to the original SIAM set used at launch time), the alignment of the PACS photometer aperture, and almost all measurements of the pointing performance were carried out with the PACS photometer point-source mode and the ``blue'' camera with the 70\,$mu$m filter (the shortest photometric wavelength) since this configuration gave the best spatial resolution. A thorough description of the PACS chopped point-source photometry observing mode is given in \cite{nielbock_this}. A catalogue of several hundred stars with strong, purely photospheric FIR flux densities compiled by the PACS Instrument Control Centre (ICC) \cite{pacs_pointing_calib_stars} was used, in order to ensure that their FIR emission was as compact as possible (i.e. no dust shells). The observations were processed using dedicated scripts within the Herschel Interactive Processing Environment (HIPE; \cite{hipe}). The sky positions of the centroids of the target stars within the 70\,$\mu$m PACS photometer maps, derived from the Herschel astrometry were compared with the catalogue positions (either Hipparcos or 2MASS). This was done for relatively large sets of targets (typically more than 40) in order to obtain distributions of offsets along the Y and Z axes. These were generally very well fitted by Gaussian distributions. The ($Y, Z$) average offsets were used to derive the alignment matrices, while the standard deviations of the distributions were used to derive a proxy, APE$^{\dagger}$\,$\equiv$\,$\sqrt{\sigma_{Y}^2 + \sigma_{Z}^2}$, for the absolute pointing accuracy that would be achieved with the alignments corrected\footnote{Beware that APE$^{\dagger}$ is not the proper APE, that is defined in temporal terms. A justification for the use of this proxy is given in \cite{craig_note}, where it is shown that it introduces an additional error of typically only 0.1\,arcsec.}.

The original set of matrices obtained from PVP observations were derived with a STR CCD reference temperature of 20$^{\circ}$C. In OD\,320, the reference temperature was lowered to -10$^{\circ}$C  in order to reduce the number of ``warm'' STR CCD pixels (see below). As a result, a bulk shift of all instruments' apertures occurred and a new SIAM set was computed accordingly. This new collection of matrices was used throughout the rest of the S/C operations. 

The main figures  of merit, namely APE, SRPE and RPE were derived from PACS photometer PVP observations (see Section \ref{subsec:launch_to_od320})\footnote{The PDE was not explicitly measured within PVP, but the long-stare pointing tests carried out indicated that the requirement was met, provided that the S/C was kept out of ``warm'' attitudes (see definition below).}. The pointing drift dependency on the Solar Aspect Angle (SAA; angle from the S/C boresight X axis to the Sun vector) was also assessed (and estimated to be significant only for SAA\,$\geq$\,110$^{\circ}$). 

In Routine Science Phase (RSP), a new  pointing calibration plan was designed \cite{rsp_plan}; its goal was twofold: on the one hand, to provide a periodic verification of the pointing calibration (alignment matrices), and on the other, to detect potential variations of the pointing performance. More than one-thousand pointing calibrations have been performed during RSP. The PACS Photometer in point-source mode (with dithering) and the ``blue camera'' with the 70\,$\mu$m filter was preferentially used. But some scan map observations were included as well (PACS Photometer mini-map mode). The pointing performance was found to be similar to that obtained in point-source mode (see below). Also, an accurate calibration of the relative misalignment of the two STR units (prime and backup) was performed in OD732 (44 calibration observations were planned and executed with the backup STR unit configured as operational. In this way, not only the misalignment of both STRs but also the pointing accuracy achievable with the backup unit was estimated). Since the accessible area of the sky was restricted during the daily telecommunications period (DTCP), RSP pointing calibration observations compatible with the DTCP restriction were generally selected. This optimised the scientific return by not blocking with these calibration measurements observing time outside the DTCP with much more flexible pointing constraints (and therefore more suitable for general scientific observations). However, the drawback of this strategy was that position-dependent effects could not be detected at earlier stages (See Section \ref{subsec:od320_to_od762} below).

One of the main pointing issues found early in the mission was the existence of the so-called ``speed bumps'', namely events of departure from the nominal attitude observed in scan maps, easily recognizable in the S/C velocity norm profile as local maxima (``bumps''). The cause of these anomalies was traced by the ACMS manufactures to the existence of a growing number of high-signal or ``warm'' pixels in the STR CCD. This problem  was corrected by lowering the reference temperature of the STR CCD to -10$^{\circ}$C in OD320. 
Unfortunately, a side effect of this action was an important increment of STR plate scale errors\footnote{Minor STR plate scale errors were present in the early phases of the mission. The result of the CCD reference temperature lowering was a substantial increase of the effect. The attitude measurements have been improved by applying corrections in the ground processing. See Table \ref{summary2} and Section \ref{subsec:implemented_improvements}.} (equivalent to unequal focal length corrections in the $Y$ and $Z$ axes), introduced by the STR CCD temperature change. These errors produced systematic, orientation-dependent offsets. The problem was partially corrected by applying a one-dimensional STR focal plane correction that took place in OD762 (this was implemented by modifying the nominal on-board focal length value of the STR, $f_0$. This is equivalent to assume identical focal length corrections in both axes, i.e. $\Delta f_Y = \Delta f_Z$). The final on-board performance was greatly improved thanks to three major developments, implemented step by step:

\begin{enumerate}
 \item In a first step (tested in OD858 and used from OD866 to OD1010), a linear two-dimensional STR correction (i.e. $\Delta f_Y \neq \Delta f_Z$) was uploaded, boosting the pointing performance to APE$^{\dagger}$\,$\simeq$\,1.0 -- 1.1\,arcsec. 
\item In a second step (since OD1011), the full STR focal plane distortion correction (8 polynomial coefficients per axis), computed in parallel by the PACS ICC and the ESA/Flight Dynamics Systems (FDS) team at the MOC was uploaded, producing a further improvement in the performance, APE$^{\dagger}$\,$\simeq$\,0.8 -- 0.9\,arcsec. 
\item In a third step (since OD1032), several ``dubious'' stars (see Section \ref{subsec:od1011_to_eol})  were removed from the on-board STR tracking catalogue, yielding the final APE$^{\dagger}$ figure rather homogenous in the sky, provided that enough tracking stars were available and that ``cold'' S/C attitude conditions applied\footnote{The term ``cold'' refers to those S/C attitudes that keep the service module away from heating by solar radiation, in contrast to ``warm'' attitudes. Originally, only those with SAA\,$\geq$\,110$^{\circ}$ were considered as problematic due to the risk of heating some components of the STR support structure.}.
\end{enumerate}

In the  following subsections, the evolution of the pointing performance is detailed. 

\subsection{Pointing performance from launch to OD320}
\label{subsec:launch_to_od320}

The most representative APE$^{\dagger}$ estimate was derived using PACS photometer point source observations during the PVP.  A large set  of observations (approximately 250) in six ODs (38, 64, 86, 92, 101 \& 104) were used to derive the  PACS photometer aperture position (named \texttt{P01\_0}) and also the absolute pointing performance figure. In addition, a large pointing campaign was implemented in OD274 to determine the pointing performance depending on the STR interlacing (il; see Section \ref{sec:acms}) mode: \textit{il-disabled}, up to 9 stars tracked at a time, and \textit{il-enabled}, from 15 to 18 stars sampled alternatively in two ``planes''. The results are summarized in Table \ref{ape_tab1}. The APE$^{\dagger}$ was measured to be between 1.9 and 2.2\,arcsec depending on the enabling of interlacing.

\begin{table}[htdp]
\caption{APE$^{\dagger}$ measurements (1$^{st}$ period). $\langle\Delta Y\rangle$ and $\langle\Delta Z\rangle$ are the average offsets of the $Y$ and $Z$ positions of the centroids of the measured stars w.r.t. the origin of the PACS photometer coordinate frame \cite{nielbock_this}. These average offsets were used to compute the SIAM of the PACS photometer {\tt P01\_0} aperture and therefore not considered for pointing error computation purposes.}
\label{ape_tab1}
\begin{tabular}{p{3cm}ccccc}
\hline\noalign{\smallskip}
OD range & $\langle\Delta Y\rangle$ & $\langle\Delta Z\rangle$ & $\sigma_Y$&$\sigma_Z$& APE$^{\dagger}$\\
& (arcsec) & (arcsec) &(arcsec) & (arcsec) & (arcsec)\\
\noalign{\smallskip}\hline\noalign{\smallskip}
38--104 (250 meas.) & -- & -- &1.09 & 1.56 & 1.90 \\
274 (all, 102 meas.) &  -0.45 & -1.71 &  1.23 &  1.51 & 1.94 \\
274 (il-disab, 55 meas.) & -0.35 & -1.67 &  1.23 &  1.72 & 2.24 \\
274 (il-enab, 47 meas.) & -0.57 & -1.75 &  0.93 &  1.22 & 1.53 \\
\noalign{\smallskip}\hline
\end{tabular}
\end{table}

\subsection{Pointing performance from OD320 to OD762}
\label{subsec:od320_to_od762}

As mentioned above, the event marking the onset of this long period is the lowering of the STR CCD reference temperature from 20 to -10$^{\circ}$C that corrected the ``speed bumps'' problem at the price of increasing the STR plate scale errors that produced systematic but boresight-dependent offsets.
Shortly after the STR CCD temperature adjustment (OD385) the first routine pointing calibration observations were executed. The plate scale errors translated into variations of the dispersion of the distribution of pointing calibration observations; nevertheless, this was not clearly and immediately reflected in the statistical results obtained from the pointing calibration campaigns: while in cycles\footnote{A cycle is a series of ODs where all the instruments are used in sequence. Typically comprises 14 ODs.} 15--22, the APE$^{\dagger}$ was consistent with PVP results, an outstanding  increase was observed in cycles 23--31 and a decrease thereafter, as summarized in  Table \ref{ape_tab2}. The investigation revealed that the effect was very small when the distribution of tracked stars was uniform across the STR FoV, but large, sometimes as much as $\sim$\,8\,arcsec when the distribution of guide stars was asymmetric (see a detailed explanation in Sect. 2.4.2 of \cite{tuttlebee_report}. This explains the different behaviour of the dispersion of measurements depending on the considered period, since, as explained above, the RSP pointing calibration observations were preferentially performed in the relatively small sky area compatible with the DTCP constraints. This sky patch changes and moves slowly across the sky and therefore varies usually  little within the time frame of a cycle. Thus, the configuration of guide stars for all pointing observations obtained within a cycle are generally similar.

\begin{table}[htdp]
\caption{APE$^{\dagger}$ measurements (2$^{nd}$ period)}
\label{ape_tab2}
\begin{tabular}{cccccccc}
\hline\noalign{\smallskip}
Cycles & OD range & No. observations & $\langle\Delta Y\rangle$ & $\langle\Delta Z\rangle$ & $\sigma_Y$&$\sigma_Z$& APE$^{\dagger}$\\
&  &  & (arcsec) & (arcsec) &(arcsec) & (arcsec) & (arcsec)\\
\noalign{\smallskip}\hline\noalign{\smallskip}
15--22 & 385--496 & 45 & -0.83 & -0.17 & 0.85 & 1.80 & 1.99\\
23--31 & 497--622 & 51 & -0.58 & -0.30 & 1.33 & 2.26 & 2.62\\
32--36 & 623--692 & 40 & -0.08 & 0.19 & 1.05 & 2.03 & 2.28\\
33--36 &  637--692 & 34 & -0.16  & 0.56 & 1.06 & 1.67 & 1.98\\
37--39 & -- &  	22 & -0.01 & -0.58 & 1.10 & 1.19 & 1.62\\
-- &  731 & 17 & 0.27 & -0.57 & 1.12 & 2.05 & 2.33\\
-- & 733 & 21 & -0.62 & 0.55 & 1.28 & 2.65 &2.95\\
\noalign{\smallskip}\hline
\end{tabular}
\end{table}

An average of a large subset of observations  in cycles 15--40 was performed (comprising nearly 200 observations). The results are shown in Table \ref{ape_tab234}. This is considered a representative value of the mean absolute accuracy achieved in this period. The largest outlier is at nearly 7\,arcsec from the barycenter of the distribution, a value that is consistent with some extreme cases reported by observers. Incidentally, the plate scale errors affected mainly the prime STR (STR1), with a negligible effect in the backup unit (STR2) due to the initial detector position\footnote{As demonstrated by measurements derived from CCD dumps performed at the initial and final STR2 CCD reference temperature.}.


\begin{table}[htdp]
\caption{average APE$^{\dagger}$ for periods 2$^{nd}$ --4$^{th}$ (see text for details)}
\label{ape_tab234}
\begin{tabular}{cccccccc}
\hline\noalign{\smallskip}
Cycles & OD range & No. observations & $\langle\Delta Y\rangle$ & $\langle\Delta Z\rangle$ & $\sigma_Y$&$\sigma_Z$& APE$^{\dagger}$\\
&  &  & (arcsec) & (arcsec) &(arcsec) & (arcsec) & (arcsec)\\
\noalign{\smallskip}\hline\noalign{\smallskip}
\multicolumn{8}{l}{2$^{nd}$ period}\\
15--40 & 385--733 & 196 & -0.40 & -0.13 & 1.17 & 2.05 & 2.36\\
\multicolumn{8}{l}{3$^{rd}$ period}\\
-- &  764 & 42 & -0.26 & 0.90 & 0.92 & 1.12 & 1.45\\
\multicolumn{8}{l}{4$^{th}$ period}\\
-- & 858 & 43 & -0.27 & 0.54 & 0.70 & 0.65 & 0.95\\
49-51 & 872--900 & 36 & -0.13 & 0.20 & 0.79 & 0.80 & 1.12\\
\noalign{\smallskip}\hline
\end{tabular}
\end{table}


\subsection{Pointing performance from  OD762 to OD866 (STR1 1D correction)}
\label{subsec:od762_to_od866}

As explained above, a first on-board corrective action was performed in OD762, by up-loading a new  nominal on-board focal length value of the STR (See Section 2.3.1 of \cite{tuttlebee_report} for details on the determination of the correction applied). The pointing performance was revisited in OD764 (42 pointing calibration observations), with very positive results, as shown in Table \ref{ape_tab234}.


\subsection{Pointing performance from OD866 to OD1010 (STR1 2D correction)}
\label{subsec:od866_to_od1010}

The linear 2D correction (i.e. two separate, linear correction factors to the Y and Z axes; see Section 2.3.1 of \cite{tuttlebee_report} for details on the determination of the correction) is the main pointing improvement carried out during the S/C operation. It was tested in DTCP858. The first assessment done by FDS was very positive. The pointing performance was verified in OD858 by means of 43 pointing calibration observations, and the results pointed towards sub-arcsec accuracy; the change was made permanent in OD866. Subsequent checks pointed towards an APE$^{\dagger}$ around 1.1\,arcsec. (Table \ref{ape_tab234})





\subsection{Pointing performance from OD1011 to the end of operations: full focal plane (FP) distortion correction}
\label{subsec:od1011_to_eol}

The full STR FP distortion correction (8 polynomial coefficients per axis) was computed by the FDS team at ESOC, based on the algorithm provided by the PACS ICC. In order to derive the final coefficients it was required to process raw data of guide stars for 20 consecutive ODs. This large volume of data was required to ensure that the parameter estimation was less sensitive to the non-calibratable biases in the measurements.  Please refer to Section 2.3.1. of \cite{tuttlebee_report} for details on the procedures applied. This correction became operational in OD1011, producing a further improvement in the performance, APE$^{\dagger}$\,$\simeq$\,0.8--0.9\,arcsec.

In a third step (since OD1032), a number (73) stars from the STR catalogue that were deemed by the PACS ICC (see \cite{aussel_stars_assess,helmut_stars_assess}) as ``dubious'' due to positional uncertainties and/or high proper motions, affecting specific pointing directions, were removed from the tracking catalogue (the ``tracking flag'' was set to zero, and thus the stars can be used for acquisition but no longer for guiding),  making the final figure (APE$^{\dagger}$\,$\simeq$\,0.8--0.9\,arcsec) rather homogeneous across the sky, assuming that a sufficient number of tracking stars are available. This value can be used as a proxy of the S/C absolute pointing accuracy during the last period of the mission operations' phase..
The pointing accuracy has nevertheless revealed quite sensitive to the SAA of the observations. Even moderately ``warm'' attitudes can produce an outstanding effect on the S/C pointing performance (see below).

\subsection{Other pointing figures}
\label{subsec:other_pointing_figures }

\subsubsection{Spatial relative pointing error (SRPE)}

The relative pointing accuracy, as measured by the Spatial Relative Pointing Error (SRPE) has been only estimated for small raster maps (nodding observations with 52\,arcsec throw). Nevertheless, even such a small size can yield representative SRPE results since the scale is larger than the usual STR bias stability range ($\sim$\,10\,arcsec). Only two estimates have been done at present: 

\begin{itemize}
\item At PVP, it was measured for non-interlaced (baseline conditions) and interlaced (goal conditions) observations, yielding an estimate of 1.54\,arcsec (baseline)/1.1 arcsec (goal)
\item In RSP, a further estimate was performed (M. Nielbock, PACS ICC) using 43 pointing calibration observations gathered in OD858 (i.e. when the STR1 2D correction was tested), yielding a result of 1.02\,arcsec (mixed interlaced and non-interlaced observations).
\end{itemize}

\subsubsection{Absolute pointing error in scan maps}

The measurements of this figure have been performed  using the PACS photometer in mini-map scan mode (two scan legs of 3\,arcmin length, with 4\,arcsec separation, at 70$^{\circ}$ and/or 110$^{\circ}$ tilt angle in instrument coordinates). Each scan direction was processed independently. The images were drizzled with a resolution of 0.1\,pixel.

An early estimate  was performed by means of 58 observations  gathered between OD320 and OD762, so it is expected that the plate scale error effect has an impact on the results. Late measurements performed in cycles 64-71 (40 observations) confirmed that, after updating the on-board STR parameters with the full focal plane distortion correction, the Herschel scan map APE$^{\dagger}$ had substantially improved; therefore, a figure of $\simeq$\,0.9\,arcsec can be considered as reference for such observations in the last period of the mission.  These figures are on the other hand almost identical to those obtained using the PACS photometer point source mode, as shown in Table \ref{ape_tab5}.

\begin{table}[htdp]
\caption{scan map APE$^{\dagger}$}
\label{ape_tab5}
\begin{tabular}{cccccccc}
\hline\noalign{\smallskip}
Cycles & OD range &  No. observations & $\langle\Delta Y\rangle$ & $\langle\Delta Z\rangle$ & $\sigma_Y$&$\sigma_Z$& APE$^{\dagger}$\\
 & &  & (arcsec) & (arcsec) &(arcsec) & (arcsec) & (arcsec)\\
\noalign{\smallskip}\hline\noalign{\smallskip}
30--33 & 595--650 & 58 & 0.38 & 0.77 & 1.24 & 2.01 & 2.36\\
64--71 & 1071--1182 & 40 & 0.61 & 0.76 & 0.52 & 0.74 & 0.90\\ 
\noalign{\smallskip}\hline
\end{tabular}
\end{table}

\subsubsection{Relative pointing error (RPE)}
\label{rpe_meas}
The relative pointing error (RPE)  has been measured during PVP, both for fixed and moving targets.  It meets or is better than  the requirement of 0.3 arcsec: RPE\,=\,0.19\,arcsec for $\alpha$ Boo and RPE\,= \,0.29\,arcsec for the asteroid 19 Fortuna.

\subsection{Solar System Objects' Tracking Performance}
\label{sso_tracking}

As stated in Section \ref{intro}, the Herschel S/C was able to point and track Solar System Objects (SSO).  These targets were uniquely defined by
specifying their  Navigation and Ancillary Information Facility (NAIF) ID\footnote{\tt http://naif.jpl.nasa.gov/naif/about.html} 
used to retrieve the information required to determine 
the coordinates at the observation time and also to calculate the
differential tracking rate required. The ephemeris information was extracted
from the NASA Jet Propulsion Laboratory (JPL) Horizons system\footnote{\tt http://ssd.jpl.nasa.gov/horizons.cgi}.
The ephemeris file for a given object contained the geometric states with
respect to the Solar System Barycentre (SSB), with a sampling period of
45 minutes. The SSO files were requested to be SSB-centric in order to perform the light-time
and aberration correction in the correct frame via an iterative Newtonian
approximation. A Hermite interpolating polynomial was used to obtain the
ephemeris between the samples.
The Herschel Scientific Mission Planning System  computed 
the spacecraft-centric geometric state combining the Horizons ephemerides, the Herschel orbit file and
JPL~DE405 planetary and lunar ephemerides\footnote{\tt http://iau-comm4.jpl.nasa.gov/de405iom/de405iom.pdf}.
Finally, offsets were applied as required
for the pointing pattern and the resulting quaternions were
converted to ACA coordinates using the
SIAM. The raster or line-scan pointings were performed relative to the SSO tracking frame. 
The tracking coefficients eventually uplinked to the S/C were defined as third-order Chebyshev polynomials describing the update quaternion as a function of time.

The SSO tracking performace was validated on two asteroids (a Jovian satellite
was also foreseen, but eventually not executed) via PACS photometer observations
executed during CoP and PVP \cite{pointing_calibration_plan,pacs_verification_plan}.
The dedicated analysis of the 70\,$\mu$m (blue) channel data (chop-nod and
scan-map observations) showed perfect point sources with point-spread function (PSF)
structures very similar to the ones obtained from fixed targets, demonstrating the
high quality of the tracking performance and fulfilling all pre-launch
requirements on tracking of moving targets. These results are consistent with the RPE measurements for moving targets performed in PVP and outlined in Section \ref{rpe_meas}. The measurements taken in the context of testing the tracking
performance (as part of the PACS focal plane geometry programme) are summarised in Table \ref{sso_obs}.

\begin{table}[htdp]
\caption{Summary of measurements performed to test the SSO tracking performance.}
\label{sso_obs}
\begin{tabular}{rcrlcrcll}
\hline
\noalign{\smallskip}
OD  & OBSID      &  \multicolumn{2}{c}{Target} & Duration & Filter &  PACS photo & S/C-centric  \\
 &  & &  &  &  & mode  & rate (arcsec/h) \\
\noalign{\smallskip}
\hline
\noalign{\smallskip}
 41 & 1342179011 &  18 & Melpomene & 3304 & blue/red & chop-nod & 67.0 \\
 41 & 1342179012 &  18 & Melpomene & 2126 & blue/red & scan-map & 67.0 \\
101 & 1342182740 &   8 & Flora     & 4954 & blue/red & chop-nod & 51.3 \\
\noalign{\smallskip}
\hline
\end{tabular}
\end{table}

In addition, tracked measurements on Mars, Neptune, Ceres, and Vesta have
been used to characterise the PACS photometer PSF \cite{pacs_psf}

Around 800 moving targets (asteroids, comets, satellites, trans-Neptunian
objects) with pre-calculated ephemeris information were available
within the Herschel Observation Planning Tool (HSpot) and 
additional targets or ephemeris updates could be requested via
Helpdesk tickets.

A few high-profile science targets required significantly faster tracking exceeding the nominal maximum speed (10\,arcsec/min):
103P/Hartley~2 had an apparent Herschel-centric speed of 30\,arcsec/min
and 45P/Honda-Mrkos-Pajdusakova moved with about 36\,arcsec/min during
the Herschel observations. In both cases the differential tracking worked
perfectly. There was only one near-Earth object which was too fast for
Herschel's tracking capabilities: 2005~YU$_{55}$ moved with 2.8-3.8 deg/hour
(168-228 arcsec/min) and crossed the entire visibility window in about
16\,hours. In this case, the observations were executed by
scanning a fixed pre-calculated field on the sky while the target
was crossing this field \cite{mueller}.

It is also worth noting the coordinate conventions in the final
data products of tracked observations. The coordinates given in the
final maps correspond to the Herschel-centric coordinates at the
{\bf start time} of the scientific measurement (the time connected to the
first data frame taken; all subsequent data frames are then``stacked''
onto this first frame), while the coordinates given in the product meta
data are refering to the object's Herschel-centric coordinates at the
{\bf mid time} of the observation.

\subsection{The impact of warm attitudes on the pointing accuracy}
\label{subsec:warm_attitudes}

\begin{figure}
\hspace{-0.5cm}
  \includegraphics[scale=0.5]{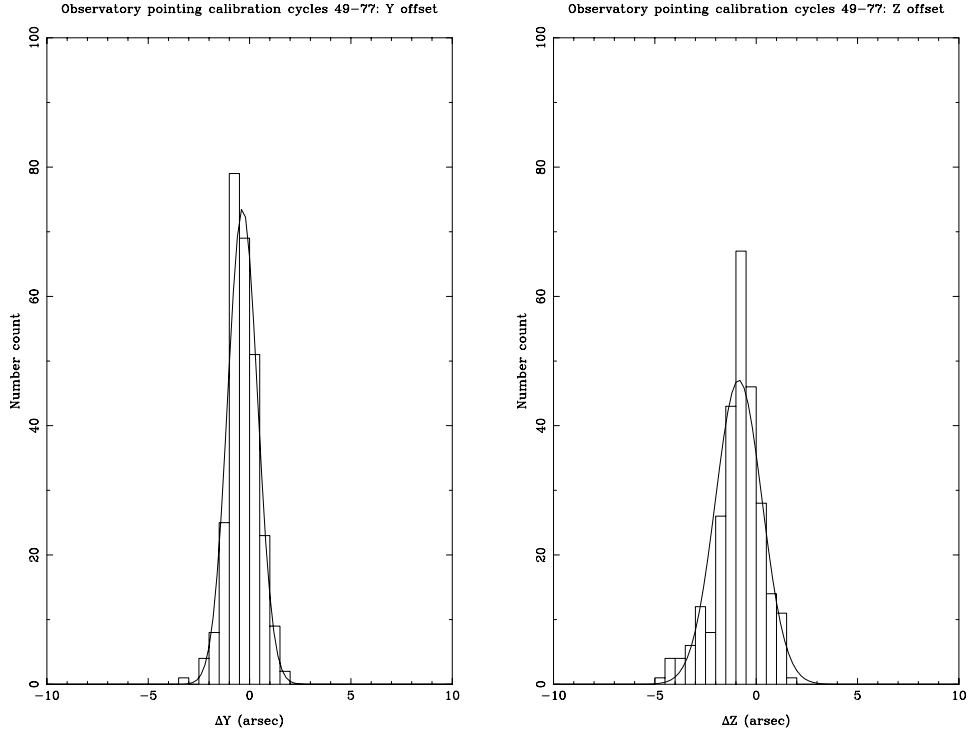}
\caption{Histograms of Y (left) and Z (right) pointing offset distributions. A clear asymmetry is observed towards the -Z direction}
\label{histo_ape_cycles_49_77}       
\end{figure}

The first RSP measurements performed after OD1032 indicated an excellent behaviour of the pointing system, i.e. APE$^{\dagger}$\,$\simeq$\,0.85\,arcsec and 0.90\,arcsec in point-source and mini-map modes, respectively. Nervetheless, a comprehensive study compiling virtually all the pointing calibration observations gathered from the major STR focal plane update (OD866; see Section \ref{subsec:od866_to_od1010}) revealed a distribution with rather large outliers, and with a clear asymmetry towards –Z  (see Fig. \ref{histo_ape_cycles_49_77}). Moreover, a relatively large offset was present in the average Y, Z position (see Table \ref{ape_tab6}).

\begin{table}[htdp]
\caption{APE$^{\dagger}$ measurements, cycles 49 to 77}
\label{ape_tab6}
\begin{tabular}{cccccccc}
\hline\noalign{\smallskip}
Cycles & OD range &  No. observations & $\langle\Delta Y\rangle$ & $\langle\Delta Z\rangle$ & $\sigma_Y$&$\sigma_Z$& APE$^{\dagger}$\\
 & &  & (arcsec) & (arcsec) &(arcsec) & (arcsec) & (arcsec)\\
\noalign{\smallskip}\hline\noalign{\smallskip}
49--77 & 872--1266 & 271 & -0.34 & -0.86 & 0.73 & 1.15 & 1.36\\
\noalign{\smallskip}\hline
\end{tabular}
\end{table}

\begin{figure}
\hspace{-0.6cm}
  \includegraphics[scale=0.9]{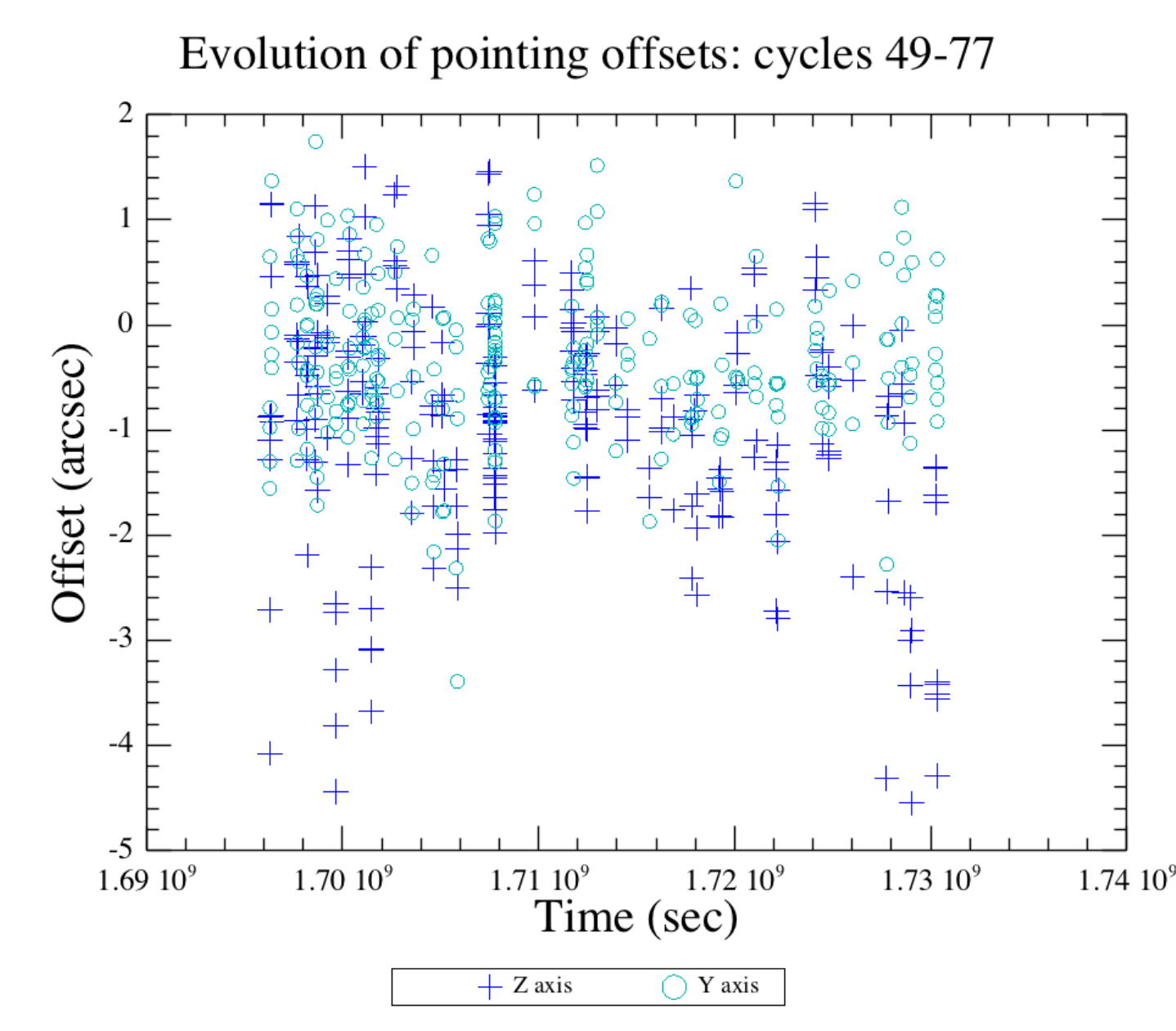}
\caption{Temporal evolution of the Y (circles) and Z (crosses) offsets from cycle 49 to 77 (OD872 to OD1266)}
\label{pointing_offset_evolution}       
\end{figure}

This indicated that an additional factor was affecting the S/C pointing. A thorough study of the evolution of  the attitude offsets across the whole period was performed (see Fig. \ref{pointing_offset_evolution}): while the Y axis showed no trend (just random dispersion around a zero offset), the Z axis exhibited a clear trend towards negative Z offsets, with two peaks separated by exactly one year. The offsets were apparently not associated to problems (e.g. incorrect astrometry) of the pointing targets themselves. Rather, the problem was traced to be related to the scheduling of long observations at negative beta (pitch) angles, corresponding to ``warm'' attitudes. As explained before, the term refers to attitudes where the STR supporting structures are subject to thermal distortions since they are not completely shaded from sunlight. In fact, in all but two cases of large astrometric Z offsets (larger than 2.5\,arcsec), long periods ($\geq$\,10\,hr) at ``warm'' attitudes (-20$^{\circ}$\,$\leq$\,$\beta$\,$\leq$\,-10$^{\circ}$ i.e. 110$^{\circ}$\,$\geq$\,SAA\,$\geq$\,100$^{\circ}$) took place two or less hours before the time of the pointing calibration observations. On the other hand, none of the correct (i.e. with reduced offset) observations investigated  (randomly picked) were carried out in or after  negative attitude periods. This indicated that the impact of relatively ``mild'' negative beta angles was larger than previously suspected. Therefore, scheduling restrictions in the range  110$^{\circ}$\,$\geq$\,SAA\,$\geq$\,100$^{\circ}$ were enforced until the end of the operational mission.

After removing several suspicious ODs (872, 911, 932, 969, 1172, 1216, 1236, 1251, 1265 \& 1266) from the statistics, the results improve, yielding a APE$^{\dagger}$ estimate of  $\approx$\,1.1\,arcsec, better aligned with previous results. One of the challenges in the pointing area is the correct parametrisation of the pointing offset dependence on the SAA. This is a complex issue, since the pointing offset depends not only on the SAA of an observation, but also on the previous values of SAA. Moreover, the time scales of the pointing drift are not properly known.

\begin{figure}
\hspace{-0.7cm}
\includegraphics[scale=0.4]{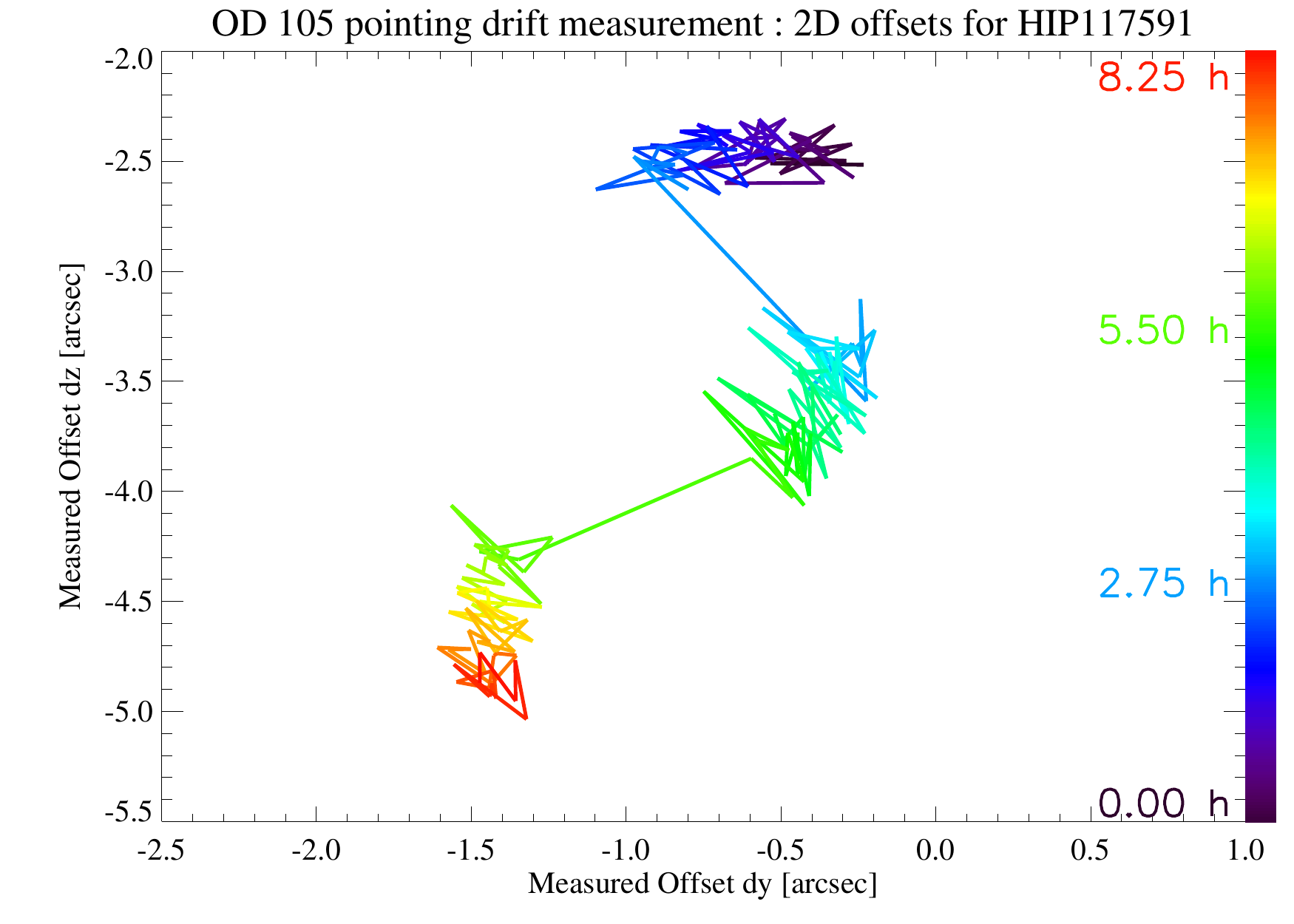}  \hspace{-0.4cm} \includegraphics[scale=0.4]{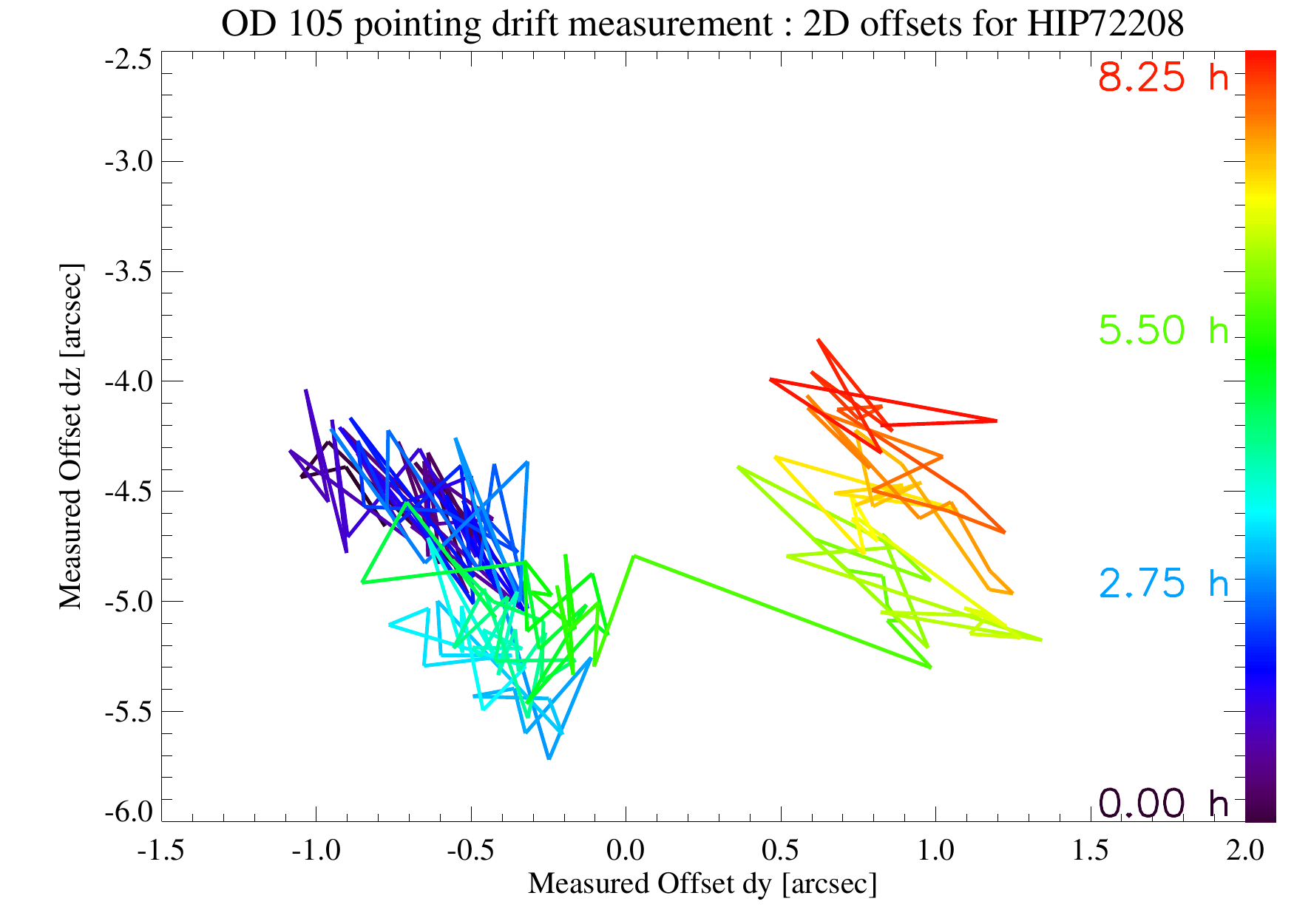}
\caption{Results of the pointing drift measurements from OD 105. Left: for star HIP~117591 (warm attitude). Right: for star HIP~72208 (cold attitude, after 12 hours of previous observations at warm attitude). The temporal evolution is indicated by the colour bar.}
\label{pointing_drifts_OD105}       
\end{figure}

One specially designed study for the pointing drift at extreme SAAs was conducted on OD 105 (August 26, 2009) within the PVP. The target HIP~117591 was observed at a SAA of 118.9 degrees. The S/C dwelled on this star for about 8.25 hours, and the centroids where monitored with the PACS photometer point-source mode observations at 70~$\mu$m. Immediately after this long staring observation (which was partitioned into three separate AORs of 2.75 hours, with their own initial target acquisition), short pointing observations were performed on a small list of stars with SAAs in the range [110,120] degrees. Finally, after around 12 hours total time of observations at these warm attitudes, the star HIP~72208 (being in the cold attitude with a SAA of 64.1 degrees) was acquired. A long staring observation for 8.25 hours, similar to the one for HIP~117591 was performed, to monitor the influence of the previous warm attitudes onto the pointing performance. We show the pointing results for the two long staring observations in Fig.~\ref{pointing_drifts_OD105}. Note that for these graphs, the previously mentioned improvements regarding the STR distortions and the revision of the STR input catalogue were not applied. The starting point in the HIP~117591 drift plot thus reflects the typical result for the pointing performance of an individual object in this early phase of the mission. The relative offsets dy and dz in arcseconds from the expected centre position are shown. Since for both stars, the measurements are separated into three parts, each having 60 chop/nod cycles, each graph contains the information of 180 elementary pointing measurements. The time evolution is indicated by a colour coding, going along the 8.25 hours per star from deep purple towards bright red. For HIP~117591, an evolution of the pointing offsets mainly in (negative) z-direction is clearly visible, an imprint of the warm attitude, which gets obvious after the first chunk of measurements, lasting for 2.75 hours. Also visible is the effect of re-pointing after 60 and 120 chop/nod cycles, which separates the offset values into three distinct clouds of points. The second graph, for HIP~72208, then demonstrates that after 12 hours at very warm attitudes, the relaxation to a normal pointing performance takes a long time. After the beginning of these observations at cold attitudes, the pointing offset in z-direction continues to increase further by roughly one arcsec, which might indicate a slight hysteresis effect in the thermal relaxation of the STR structures. Even after dwelling for 8.25 hours at this cold attitude, the pointing offset in z-direction did not improved much compared to the beginning of the measurement. This demonstrated that such extreme warm attitudes had to be avoided for normal S/C operations. It further showed that the temporal evolution of the pointing degradation, and an eventual relaxation to normal performance was a quite complex effect. 

\section{Improving the pointing performance}
\label{sec:pointing_improvement}

This section describes the efforts already implemented, currently on-going and future developments aimed towards optimising the \textit{a posteriori} knowledge of the S/C pointing and therefore the accuracy of the astrometry in our products.

\subsection{Implemented improvements}
\label{subsec:implemented_improvements}

A simple offset correction software has been implemented at the HSC, applying the STR focal plane 2D linear correction (Section 2.4.1 of \cite{tuttlebee_report}) to each attitude within the pointing product.  In summary, the 
software logic applies the steps below:
 \begin{enumerate}
\item Load catalogue of stars from the STR, furnished by the ACMS manufacturers.
\item Open the pointing product and read its records in a loop. For each attitude sample:
 \begin{enumerate}
   \item Get the filtered attitude and the number of stars tracked (furnished in the STR quality flag field).
   \item Get the catalogue information from those stars within the FoV of the STR at the current S/C filtered 
       attitude\footnote{The STR is roughly pointing in the opposite direction to the telescope boresight.}: X,Y,Z, magnitude, and trackability status.
    \item For the trackable stars within the FoV, determine the direction vectors using the
       STR ``focal length'' linear correction factors provided by FDS. These linear factors are introduced in the following way: the direction vector of a star in the STR boresight frame can be expressed in the small angle approximation as:
       \begin{equation}
       \mathbf{u_s} =  \left(
        \begin{array}{c}
        1\\
        -y/f\\
        \ z/f\\ 
        \end{array}\right) \sqrt{1 + (y/f)^2 + (z/f)^2}
       \end{equation}
       \noindent Where $y$ and $z$ are the star's coordinates in the STR CCD frame, and $f$ is the STR telescope's focal length. The linear ``focal length'' correction terms are introduced by replacing $f$ by $f + \Delta f_y$ and $f + \Delta f_z$ in the Y and Z axes, respectively. Therefore, $y/f$ gets replaced by  $ (y / f) ( 1 - (\Delta f_y / f)  +   O((\Delta f_y / f)^2)) $ and an identical expression for the Z axis. Then, the corrected star vectors can be approximated as:
       \begin{equation}
       \mathbf{u'_s} \approx  \left(
        \begin{array}{c}
        1\\
        -\frac{y}{f} \left( 1 - \frac{\Delta f_y}{f} \right) \\
        \ \frac{z}{f} \left( 1 - \frac{\Delta f_z}{f} \right) \\
        \end{array}\right) \sqrt{ 1 + \left[ \frac{y}{f} \left( 1 - \frac{\Delta f_y}{f} \right) \right]^2 + \left[ \frac{z}{f} \left( 1 - \frac{\Delta f_z}{f} \right) \right]^2}
       \end{equation}

   \noindent  Specific correction factors have been derived and provided by FDS for the different mission periods up to OD866 (when the linear 2D was implemented on-board).
                  
   \item Determine the tracked stars by means of the same selection algorithm used on-board the STR (limited to a maximum         
   of nine stars)
   \item For the chosen  stars, given the direction vectors and the inertial
      (reference) vectors, determine the best attitude using the \textit{\textbf{q} Method}  (see \cite{Wertz} and references therein) that provides a means for computing an optimal three-axis attitude from many vector observations. 
   \item  Replace the filtered quaternion by the modified quaternion computed by the \textit{\textbf{q} Method}.
 \end{enumerate}
\end{enumerate}

The applied correction does not add noise to the attitude samples since it is a purely geometric one. When the correction factors are set to 1.0, the input attitude is exactly recovered (within $\approx$\,1.0\,$\times$\,10$^{-10}$\,arcsec). In the current implementation, the software has some limitations, namely: {\em (i)} STR interlacing is not considered (a single STR ``plane'', using up to nine stars is processed). {\em (ii)} The stars' selection algorithm uses the same logic as the STR. But there is no check that the stars used within the script are the same as those actually used for tracking. In particular, if one of the 73 bad stars was used for tracking, it will be replaced by a different one by the correction software. {\em (iii)}  The 2D linear correction is implemented, rather than the full one that includes eight polynomial terms per axis.

\begin{figure}
\hspace{-0.5cm}
  \includegraphics[scale=0.5]{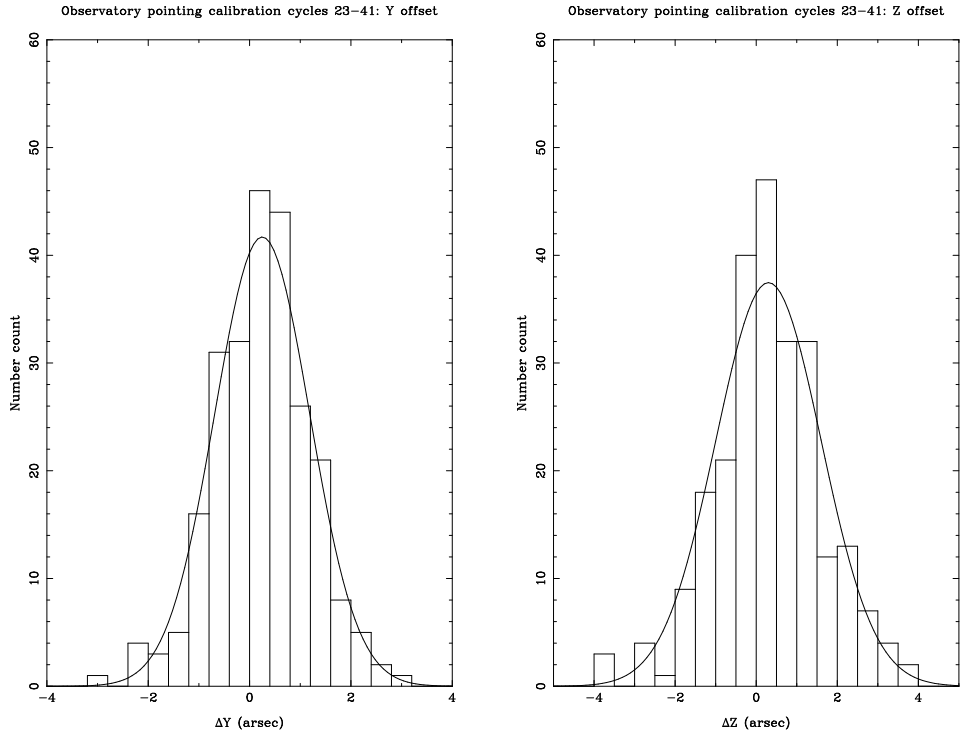}
\caption{Histograms of Y (left) and Z (right) pointing offset distributions obtained after applying the pointing reconstruction algorithm described in Section \ref{subsec:implemented_improvements}.}
\label{histo_ape_reconstructed}       
\end{figure}

\begin{figure}
  \includegraphics[scale=0.42]{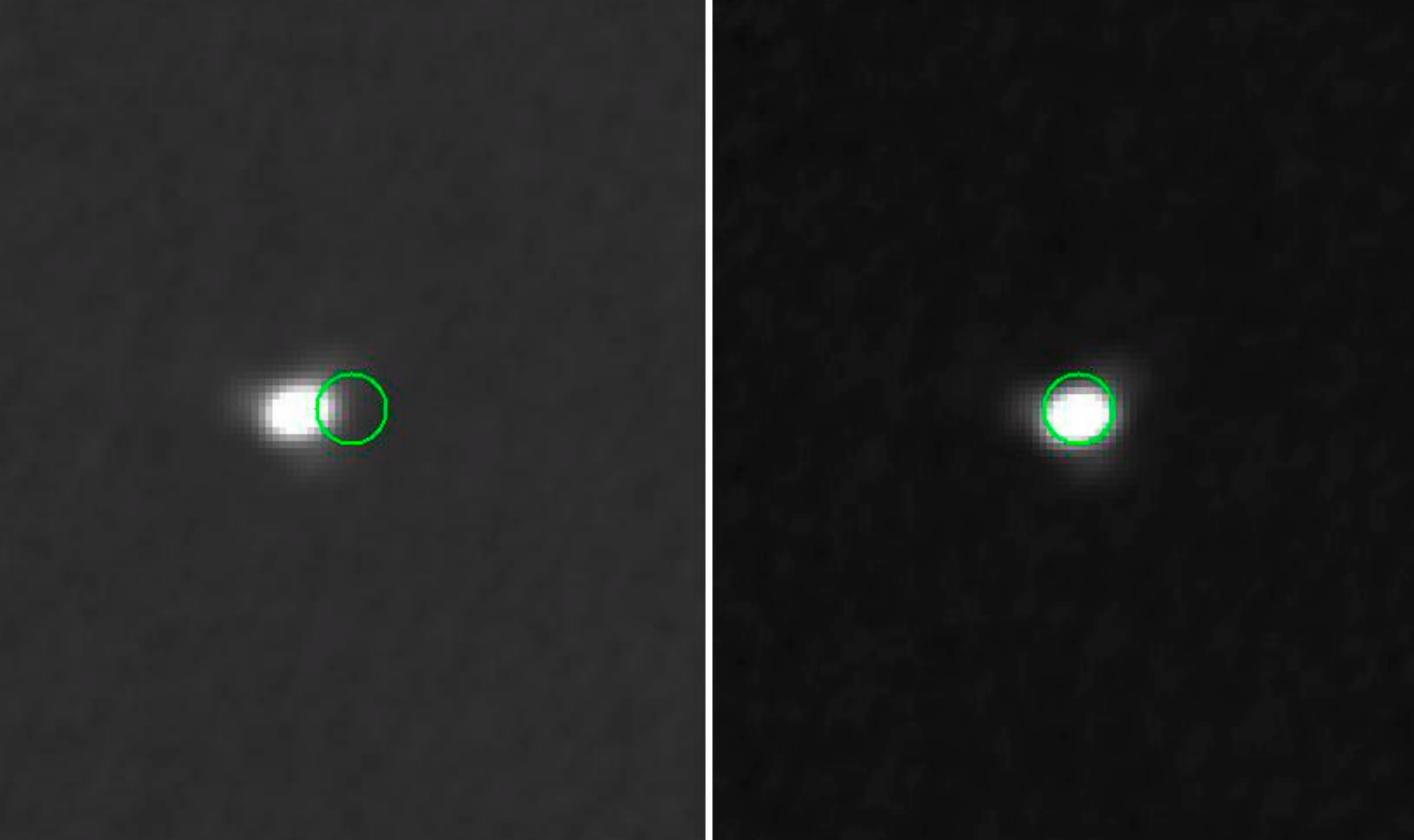}
\caption{A dramatic example of the effects of the simple pointing reconstruction described in  Section \ref{subsec:implemented_improvements}: The PACS 70\,$\mu$m map of the star UZ Tau was obtained in OD684. The observations towards the Taurus region show in general unusually large offsets due to the very asymmetric configuration of trackable stars in the STR boresight. The green circles are centred on the astrometric catalogue position of the target. The left panel shows the map with the original pointing product, while the right one presents the map processed with the reconstructed pointing product.The absolute astrometry offset is reduced from 7.5\,arcsec to less than 1\,arcsec.}
\label{pointing_offset_UZ_Tau}       
\end{figure}

The simple reconstruction algorithm described above has been thoroughly tested in different scenarios. In particular, the full set of pointing calibration observations performed in cycles 23 to 41, comprising 245 observations has been reprocessed using the improved pointing products. The results are shown in Fig. \ref{histo_ape_reconstructed} and in Table \ref{ape_reconstructed}. These results can be compared with those shown in Table \ref{ape_tab234}. An outstanding example of the improvements achievable in some extreme cases is shown in Fig. \ref{pointing_offset_UZ_Tau}. At the time of writing this paper, this correction has been applied to all observations in the Herschel Science Archive (HSA) from OD320 to OD865  (in OD866 the STR1 2D correction was applied and this ground correction is therefore not required) and a correction has been recently implemented for the first period of the mission (before OD320) and and will be applied to the observations in the HSA in the next bulk reprocessing.  A summary of the accuracies  achieved with the attitude data as given by the S/C and after ground processing is given in Table \ref{summary2}.

\begin{table}[htdp]
\caption{APE measurements using the reconstructed pointing products.}
\label{ape_reconstructed}
\begin{tabular}{cccccccc}
\hline\noalign{\smallskip}
Cycles & OD range &  No. observations & $\langle\Delta Y\rangle$ & $\langle\Delta Z\rangle$ & $\sigma_Y$&$\sigma_Z$& APE$^{\dagger}$\\
 & &  & (arcsec) & (arcsec) &(arcsec) & (arcsec) & (arcsec)\\
\noalign{\smallskip}\hline\noalign{\smallskip}
23--41 & 497--762 & 245 & 0.25 & 0.30 & 0.94 & 1.30 & 1.60\\
\noalign{\smallskip}\hline
\end{tabular}
\end{table}

\subsection{On-going developments and planned improvements}
\label{subsec:planned_improvements}

There are a number of on-going and planned developments to improve the \textit{a posteriori} knowledge of the S/C pointing. A large effort has been done by the PACS ICC to improve the attitude measurements given by the STR, based not only on a full focal plane distortion correction but also taking into account the effects of the STR sub-pixel effects on the star centroiding caused by the pixel insensitive borders \cite{helmut_paper}. In addition, algorithms to propagate the attitude using the rate measurements given by the gyroscopes by means of the kinematic equation of motion (see for instance \cite{Wertz}, p. 511) have been developed and benefit from the improved STR measurements (better estimates of the gyroscopes' drift rate biases and more accurate absolute pointing directions). Several studies have indicated that the pointing jitter may be accurately reconstructed using gyroscopes' measurements alone \cite{aussel11,aussel12,helmut12b,tuttlebee_report}.

In addition, several corrections affecting specific observations or ODs are being incorporated into the standard pipeline processing (SPG version 12.x and later). These include corrections for switch-over events from the prime to the redundant STR (these introduce offsets of some 15\,arcsec), corrections for resets of the S/C velocity vector (SVV) values used by the STR on-board software to compute the aberration correction of the coordinates of the guide stars (the lack of this correction can introduce offsets up to $\approx$\,20\,arcsec) and \textit{ad-hoc} corrections for pointing offsets (sometimes of several arcsec) induced by extreme ``warm'' attitudes. 

Future developments being studied (to be implemented in the time frame 2014--2016) will likely include an attitude improvement software using batch least-squares estimators (these estimators update the state vector at a time using a block of observations taken during a fixed time span, as opposed to sequential estimators -e.g. Kalman filters- that update the state vector after each observation; see for instance \cite{Wertz}, pp. 448-459) and a general modelling of the pointing offset induced by thermo-elastic distortions within the S/C.

\section{Summary and conclusions}
\label{conclusions}
The Herschel's pointing system has experienced a complex evolution across the operational life  of the S/C. Tables \ref{summary1} and \ref{summary2} summarise the sequence of main events and the evolution of the pointing accuracy accross the operational life of the observatory. The PVP demonstrated that the procedures devised within the calibration plan \cite{pointing_calibration_plan} were adequate to accomplish the mission of calibrating the alignment of a set of FIR and sub-mm instruments w.r.t. the S/C platform, without the help of ancillary optical instruments (e.g. the Quadrant Star Sensor on-board the Infrared Space Observatory --ISO; \cite{iso}). Moreover, the ACMS actual performance was found within the requirements and pre-flight predictions (only a non-compliance with the SRPE requirement was anticipated). Nevertheless, the reduction of the STR CCD reference temperature that was very successful to solve the ``speed bumps'' problem brought an unforeseen side-effect due to the plate scale changes within the STR. This new problem was hard to detect due to its dependence on the distribution of guide stars within the STR FoV, and highlights the importance of a continuous monitoring of the S/C pointing performance (preferentially with a wide sky coverage) to cope with changes in the ACMS behaviour. On the other hand, the pointing accuracy was found more sensitive than anticipated to attitudes where the STR supporting structures are subject to thermal distortions. This was also revealed thanks to large number of routine pointing calibration checks. In any case, it is worth stressing the fact that, even in the worst circumstances, the performance was within the requirements set (but for the SRPE as mentioned above). However, given a clear scientific motivation for further improvement an outstanding effort has been done and is still being invested in order to get the best from the attitude information. This collaborative effort is shared by different actors (at MOC, HSC, ICCs and industry) and reveals the high value of the communication and cooperation between different teams within the ground segment. 

\begin{table}[htdp]
\caption{Timeline of events}
\label{summary1}
\begin{tabular}{cp{4.5cm}p{6cm}}
\hline\noalign{\smallskip}
OD & Event  & Effect\\
\noalign{\smallskip}\hline\noalign{\smallskip}
320 & Lowering STR CCD reference temperature & ``speed bumps'' disappeared but STR plate scale errors suffered a considerable increment.\\
762 & STR 1D correction & Improvement of the S/C-provided attitude information up to APE$^{\dagger}$\,$\approx$\,1.4\,arcsec\\
866 & STR 2D correction & Further improvement of accuracy up to APE$^{\dagger}$\,$\approx$\,1.1\,arcsec\\
1011 & STR full FP correction & Further improvement of accuracy up to APE$^{\dagger}$\,$\approx$\,0.9\,arcsec\\
1032 & STR catalogue update (cleanup) & Improved accuracy in specific areas of the sky\\
\noalign{\smallskip}\hline
\end{tabular}
\end{table}

\begin{table}[htdp]
\begin{minipage}{\textwidth}
\caption{Summary of the evolution of the Herschel's astrometrical accuracy.}
\label{summary2}
\begin{tabular}{lccl}
\hline\noalign{\smallskip}
OD range & Raw accuracy & Ground-processed  & Ground-correction\\
& APE$^{\dagger}$ (arcsec) & APE$^{\dagger}$ (arcsec) & available in SPG\\
\noalign{\smallskip}\hline\noalign{\smallskip}
32--320  & 1.9--2.2 & 1.4 & $>$\,v11.1.0 \\
321--761 & 2.4\footnote{extreme outliers at $\ge$\,8 arcsec possible} & 1.6 &  $>$\,v9.1.0   \\
762--865 & 1.45 & 1.3 &  $>$\,v10.0.3  \\
866--1010 & 1.1 & -- &  N/A \\
1011--1452 (EoH) & 0.9 & -- &  N/A \\  
\noalign{\smallskip}\hline
\end{tabular}

\end{minipage}

\end{table}

\begin{acknowledgements}

We acknowledge the anonymous referee for his/her valuable comments and suggestions.
The Herschel spacecraft was designed, built, tested, and launched under a contract to ESA managed by the Herschel/Planck Project team by an industrial consortium under the overall responsibility of the prime contractor Thales Alenia Space (Cannes), and including Astrium (Friedrichshafen) responsible for the payload module and for system testing at spacecraft level, Thales Alenia Space (Turin) responsible for the service module, and Astrium (Toulouse) responsible for the telescope, with in excess of a hundred subcontractors.

PACS has been developed by a consortium of institutes led by MPE (Germany) and including UVIE (Austria); KU Leuven, CSL, IMEC (Belgium); CEA, LAM (France); MPIA (Germany); INAF-IFSI/OAA/OAP/OAT, LENS, SISSA (Italy); IAC (Spain). This development has been supported by the funding agencies BMVIT (Austria), ESA-PRODEX (Belgium), CEA/CNES (France), DLR (Germany), ASI/INAF (Italy), and CICYT/MCYT (Spain).
\end{acknowledgements}

\begin{table}[htb]
\begin{minipage}{\textwidth}
\caption{Contents of the pointing product (columns of each table data set/binary table)}
\label{pointingProd}
\begin{tabular}{p{2cm}lllp{5cm}}
\hline\noalign{\smallskip}
Description & Name & Format & Unit & Comment \\
\noalign{\smallskip}\hline\noalign{\smallskip}
On-board time & obt & Double1d & $\mu$s & \\
Commanded pointing quaternion & commandQuat & Double2d & -- & Contains the four double-precission 
components of the commanded attitude quaternion\footnote{The order of the components in any quaternion array is 
such as the scalar component goes in the last position.} in the ACA frame.\\
Filtered attitude quaternion & filterQuat &  Double2d & -- & Contains the four double-precission 
components of the filtered attitude quaternion in the ACA frame.\footnote{The ``filtered'' attitude 
refers to the attitude recomputed on-ground based on the available improvement
algorithms, if available. Otherwise the column contains the same values as those stored in
the ``uncorrected filtered'' attitude.}\\
Gyro-propagated attitude quaternion & gyroPropQuat & Double2d & -- & Contains the four double-precission 
components of the gyro-propagated attitude quaternion in the ACA frame.  If on-ground gyro-propagation is not enabled or available for the data record, this field contains a ground-derived filtered quaternion based on the nominal estimator.\footnote{The ``gyro propagated'' attitude refers to the attitude derived using the GYR outputs
in telemetry to propagate an attitude quaternion derived from the on-board filtered
attitude quaternion at an OFF-position prior to the execution of a raster
or line scan.}\\
STR quality index & strQuality & Double1d & arcsec & Attitude quality index based on STR data. This quality index can be used as an error measurement for the filtered attitude.\\
Gyro-propagated quality index & gyroQuality & Double1d & arcsec & A quality index based on gyro propagation. If on-ground gyro-propagation is not enabled or available for the data record, this field contains a copy of the STR quality index.\\
S/C angular velocity & angVelocity & Double2d & arcsec/s &  \\
S/C angular velocity error & angVelocityErr &  Double2d & arcsec/s & \\
Constant velocity flag & isConstantVelocity & Bool1d & -- & Relevant to scan maps. Set to 1 when the scan speed gets constant. \\
Quality flag & qualityFlag & Int1d & -- & Its value corresponds to the number of tracked stars by the STR (the higher the number the better the quality of the derived attitude).\\
STR interlacing status & isInterlacing & Bool1d & -- & 1 if STR interlacing active, 0 otherwise.\\
Slew flag & isSlew (ISSLEW) & Bool1d & -- & Set to 1 when the S/C is slewing.\\
On-target flag &  isOnTarget & Bool1d  & -- & Set to 1 when the S/C is on-target.\\
Off-position flag &  isOffPosition & Bool1d & -- & Set to 1 when the S/C is at the attitude reference position (OFF position).\\
Out-of-field flag & isOutOfField & Bool1d & -- & Set to 1 when the S/C is at the out-of-field referece attitude.\\
Uncorrected filtered attitude quaternion &  uncorrFilterQuat &  Double2d & -- &  Contains the four double-precission components of the STR-uncorrected filtered attitude quaternion in the ACA frame.\footnote{The ``uncorrected filtered'' attitude 
refers to the attitude computed by the ACMS combining STR and GYR outputs and down-linked in telemetry.}\\
Solar Aspect Angle & solarAspectAngle &  Double1d & deg & Solar Aspect Angle (SAA) as provided in the AHF.\\
\noalign{\smallskip}\hline
\end{tabular}

\end{minipage}

\end{table}



\end{document}